\documentclass[prb,superscriptaddress,onecolumn,preprint,tightenlines]{revtex4}
\usepackage{amsmath}
\usepackage{amssymb}
\usepackage[dvips]{graphicx}

\input{tcilatex}

\begin{document}

\title{Cluster pair correlation function of simple fluids: energetic
connectivity criteria}
\author{Luis A. Pugnaloni}
\email{luis@iflysib.unlp.edu.ar}
\affiliation{Instituto de F\'{\i}sica de L\'{\i}quidos y Sistemas Biol\'{o}gicos
(IFLYSIB), UNLP-CONICET, cc. 565, (1900) La Plata, Argentina}
\author{Guillermo J. Zarragoicoechea}
\affiliation{Instituto de F\'{\i}sica de L\'{\i}quidos y Sistemas Biol\'{o}gicos
(IFLYSIB), UNLP-CONICET, cc. 565, (1900) La Plata, Argentina}
\affiliation{Comisi\'{o}n de Investigaciones Cient\'{\i}ficas y Tecnol\'{o}gicas de la
Prov. de Buenos Aires (CICPBA), Argentina}
\author{Fernando Vericat}
\affiliation{Instituto de F\'{\i}sica de L\'{\i}quidos y Sistemas Biol\'{o}gicos
(IFLYSIB), UNLP-CONICET, cc. 565, (1900) La Plata, Argentina}
\affiliation{Grupo de Aplicaciones Matem\'{a}ticas y Estad\'{\i}sticas de la Facultad de
Ingenier\'{\i}a (GAMEFI) - UNLP, La Plata, Argentina}

\begin{abstract}
We consider the clustering of Lennard-Jones particles by using an energetic
connectivity criterion proposed long ago by T.L. Hill [J. Chem. Phys. 
\textbf{32}, 617 (1955)] for the bond between pairs of particles. The
criterion establishes that two particles are bonded (directly connected) if
their relative kinetic energy is less than minus their relative potential
energy. Thus, in general, it depends on the direction as well as on the
magnitude of the velocities and positions of the particles. An integral
equation for the pair connectedness function, proposed by two of the authors
[Phys Rev. E \textbf{61}, R6067 (2000)], is solved for this criterion and
the results are compared with those obtained from molecular dynamics
simulations and from a connectedness Percus-Yevick like integral equation
for a velocity-averaged version of Hill's energetic criterion.
\end{abstract}

\maketitle

\section{Introduction}

The concepts of clustering and percolation have been widely used in order to
explain several phenomena in very diverse areas including Physics,
Chemistry, Biology, Geology, Sociology and Economics. In particular, with
reference to chemical--physics, phenomena such as nucleation,\cite{Senger1}
hydrogen bonding,\cite{Starr1} insulator--conductor, sol--gel and glass
transitions \cite{Simon1,Chen1,Coniglio1,Butler1,Stanley1,Grest1,Wittmann1}
as well as bridging in granular materials \cite{Pugnaloni1} are currently
studied from this point of view. In all these cases, the system under study
can be thought of as a collection of individuals (atoms, molecules, grains,
etc.) that, with generality, we call particles. Most of the efforts have
been based on lattice representations of the systems. The relative
simplicity of lattice models allows for a wide variety of treatments, which
extend from almost heuristic \cite{Sahimi2} to quite rigorous.\cite%
{Grimmett1} Whatever the treatment is, the concept of connectivity between
the particles plays an important role.

Sometimes, however, a continuous description---where particles can occupy
any point in a continuum phase space---is needed to reach a more realistic
picture of the phenomena under consideration. For this context, the concept
of connectivity has been generalized and adapted to describe clustering and
percolation in continuum systems. The main ideas have been established in
the pioneering works of Hill \cite{Hill1} and Coniglio \textit{et al.}\cite%
{Coniglio2} Hill considers a partition of the whole system into subsystems
of particles (the clusters) that satisfy some linking properties. The
concept of cluster is thus directly related to the idea of bonded pairs. A
bonded pair is a set of two particles that are linked by some direct
mechanism. A cluster is then defined as a set of particles such that any
pair of particles in the set is connected through a path of bonded pairs. We
call these clusters \textquotedblleft chemical clusters\textquotedblright\
to distinguish them from the non-pair-bonded clusters we have introduced in
a previous work \cite{Pugnaloni2}---note, however, that this does not mean
that clusters are necessarily formed through a true chemical bonding. A
system is said to be in a percolated configuration if it contains a cluster
that spans the system volume.

From Hill's theory, we see that a connectivity criterion is needed in order
to decide whether two particles are bonded or not. This connectivity
criterion has to be defined in accordance with the phenomenon under study. 
\cite{comment1,Chen1,Pugnaloni3} In the search for stable atomic clusters,
which mark the onset of a phase transition in a monatomic gas, Hill proposed
a simple energetic criterion (HE): two particles are bonded if their
relative kinetic energy is less than the negative of their relative
potential energy.\cite{Hill1} In principle, this criterion takes into
account the relative positions and velocities of the relevant pair of
particles. For molecular fluids, instead of just their distance, the
potential energy could in general depends on the direction and magnitude of
the vector position of each of the two involved particles and their relative
orientations.

From a theoretical point of view, a criterion that involves the velocity of
the particles prevents the straightforward integration of the momenta in the
partition function, which is the great advantage of classical statistical
mechanics. To avoid this obstacle, Hill \cite{Hill1} himself has proposed a
velocity-averaged (VA) version of his criterion giving effective potentials
between bound and unbound particles.

The VA and the complete HE criteria have been used in computer simulations
as well as in integral equations studies. For the VA criterion only the
particle positions come into account, so it is suitable to both, Monte Carlo
(MC) and molecular dynamics (MD) calculations. With respect to the integral
equations approach, Coniglio \textit{et al.}\cite{Coniglio2} have obtained a
connectedness Ornstein--Zernike (OZ) relationship for the pair connectedness
function $g^{\dagger }(\mathbf{r}_{1},\mathbf{r}_{2})$ (see also a review
from Stell).\cite{Stell1} This function is proportional to the joint
probability density of finding two particles belonging to the same cluster
and at positions $\mathbf{r}_{1}$ and $\mathbf{r}_{2}$, respectively.
Therefore, by integrating $g^{\dagger }(\mathbf{r}_{1},\mathbf{r}_{2})$, the
mean cluster size $S$ and the percolation density $\rho _{p}$---\textit{i.e.}
the value of $\rho $ for which $S(\rho )$ diverges---can be obtained. Since
Coniglio's theory deals only with the positions of the particles, the HE
criterion cannot be implemented. Instead, the VA criterion was used by
Coniglio \textit{et al.} \cite{Coniglio2} to analytically calculate the
percolation loci, for a potential made up of a hard core plus an attractive
interaction, in a crude mean-field approximation.

It is worth mentioning that most of the theoretical studies\cite%
{Chiew1,DeSimone1,Laria1,Carlevaro1} on connectivity and percolation in
continuum systems based on Coniglio's type equations were focused in the
rather simple Stillinger's connectivity criterion. \cite{Stillinger1} This
criterion states that two particles are bonded if they are separated by a
distance shorter than a given connectivity distance $d$. In this case, $d$
is an \textit{ad hoc} parameter, which must be chosen on physical grounds.
Although this criterion might be sensible in the study of certain
insulator--conductor transitions, it is unrealistic regarding clustering in
saturated vapors.

A general theory which is appropriate for bonding criteria involving both,
the momenta and positions of a pair of particles, has been developed by two
of us. \cite{Pugnaloni4} The main object in our theory is the pair
connectedness function $g^{\dagger }(\mathbf{r}_{1},\mathbf{r}_{2},\mathbf{p}%
_{1},\mathbf{p}_{2})$ which is proportional to the joint probability density
of finding two particles at positions $\mathbf{r}_{1}$ and $\mathbf{r}_{2}$
with momenta $\mathbf{p}_{1}$ and $\mathbf{p}_{2}$, respectively, and
belonging to the same cluster. This function verifies also an OZ like
relationship. In a previous paper \cite{Pugnaloni5} (thereafter denoted as
I) we applied our general theory to study the complete HE criterion for the
same model fluid considered by Coniglio \textit{et al.} \cite{Coniglio2}
under the VA criterion. We also used the same simple closure relation
proposed by Coniglio \textit{et al. }More recently, we have reported \cite%
{Zarragoicoechea1} the solution of our generalized connectedness OZ type
relation for a Lennard--Jones fluid closed with a connectedness
Percus--Yevick (PY) condition. We implemented a connectivity criterion which
generalizes Stillinger's criterion\ in that a life time $\tau $ for the
bonds is required. \cite{Pugnaloni2}

In Ref. I we have also performed MD simulations of the Lennard--Jones fluid
and have used both criteria (HE and VA) \ to define the clusters. We
concluded that the VA criterion strongly overestimates percolation
densities. We will partially revise these results here and will discuss some
subtleties related to the identification of percolating clusters. Notice
that MD simulations are convenient when the HE criterion is used to identify
clusters since MC algorithms do not provide \textit{per se} the velocities
of the particles. \cite{comment2} It should be mentioned that the HE
criterion has been considered in MD studies of small clusters and the
critical percolation behavior of Lennard--Jones fluids by several authors. 
\cite{Soto1,Soto2,Campi1} It has been suggested that the percolation
line---the line that separates the temperature--density phase diagram into
percolated and non-percolated states---might be experimentally observable. 
\cite{Campi1,Coniglio3} Moreover, cluster analysis based on this criterion
seems to be useful in locating the gas--liquid coexistence curve. \cite%
{Campi1}

The main purpose of this work is to apply the generalized connectedness OZ
type relationship closed with a connectedness PY condition for the
Lennard--Jones fluid in order to obtain the pair connectedness function $g_{%
\text{HE}}^{\dagger }(\mathbf{r}_{1},\mathbf{r}_{2},\mathbf{p}_{1},\mathbf{p}%
_{2})$ for HE clusters and thus the related cluster pair correlation
function:

\begin{equation}
g_{\text{HE}}^{\dagger}(\mathbf{r}_{1},\mathbf{r}_{2})=\int\rho(\mathbf{r}%
_{1},\mathbf{p}_{1})\rho(\mathbf{r}_{2},\mathbf{p}_{2})g_{\text{HE}%
}^{\dagger }(\mathbf{r}_{1},\mathbf{r}_{2},\mathbf{p}_{1},\mathbf{p}_{2})d%
\mathbf{p}_{1}d\mathbf{p}_{2},  \label{1}
\end{equation}
where $\rho(\mathbf{r}_{1},\mathbf{p}_{1})$ is $N$ times the probability
density of finding a particle at the phase space configuration $(\mathbf{r}%
_{1}$, $\mathbf{p}_{1})$.

We compare $g_{\text{HE}}^{\dagger }(\mathbf{r}_{1},\mathbf{r}_{2})$ so
obtained with the function $g_{\text{VA}}^{\dagger }(\mathbf{r}_{1},\mathbf{r%
}_{2})$ for the VA criterion calculated using the integral equation that
results when the OZ type relationship of Coniglio \textit{et al. }\cite%
{Coniglio2} is closed with a PY like conditions. Both functions---$g_{\text{%
HE}}^{\dagger }(\mathbf{r}_{1},\mathbf{r}_{2})$ and $g_{\text{VA}}^{\dagger
}(\mathbf{r}_{1},\mathbf{r}_{2})$---are compared with the corresponding
curves given by MD simulations.

The paper is organized as follow. In Sec. II we present the model system and
the two connectivity criteria, i.e. HE and VA, we will work on. Also we use
this section to discuss some aspects about the MD simulations. The continuum
clustering theories suitable for each criteria will be sketched in Sec. III.
There, we briefly describe the integral equation for $g_{\text{HE}}^{\dagger
}(\mathbf{r}_{1},\mathbf{r}_{2},\mathbf{p}_{1},\mathbf{p}_{2})$ and its
solution following Lado's orthogonal polynomials method. \cite%
{Lado1,Zarragoicoechea1} Finally in Sec. IV we compare the theoretical
results with those obtained from simulations. We then summarize and give the
conclusions.

\section{Model system and energetic criteria}

We consider a system of $N$ particles whose configurations are given by
their positions and momenta $(\mathbf{r}_{i},\mathbf{p}_{i})$ $($\textit{i = 
}$1,...,N)$. The canonical ($NVT$) ensemble will be used throughout. We
assume that particles interact via the Lennard--Jones pair potential

\begin{equation}
v(r_{ij})=4\varepsilon\left[ \left( \frac{\sigma}{r_{ij}}\right)
^{12}-\left( \frac{\sigma}{r_{ij}}\right) ^{6}\right] ,  \label{2}
\end{equation}
where $r_{ij}=\left\vert \mathbf{r}_{ij}\right\vert $ with $\mathbf{r}_{ij}=%
\mathbf{r}_{j}-\mathbf{r}_{i}$.

The clustering criteria are expressed in terms of the bond conditional
probability density $P(\mathbf{r}_{i,j},\mathbf{p}_{i,j})$, say the
probability density that two particles $i$ and $j$ are bonded under the
condition that their positions and momenta are $(\mathbf{r}_{i},\mathbf{p}%
_{i})$ and $(\mathbf{r}_{j},\mathbf{p}_{j})$, respectively.

\subsection{HE criterion}

The original Hill's criterion (HE) identifies clusters by defining: \cite%
{Hill1}

\begin{equation}
P_{\text{HE}}(\mathbf{r}_{i,j},\mathbf{p}_{i,j})=\left\{ 
\begin{array}{ll}
1 & \mathbf{p}_{i,j}^{2}/4m<-v(\mathbf{r}_{i},\mathbf{r}_{j})\text{ \ and \ }%
r_{i,j}\leq d \\ 
0 & \mathbf{p}_{i,j}^{2}/4m\geqslant -v(\mathbf{r}_{i},\mathbf{r}_{j})\text{
\ or \ }r_{i,j}>d%
\end{array}%
\right.  \label{3}
\end{equation}%
with $\mathbf{p}_{i,j}$ the relative momentum: $\mathbf{p}_{i,j}=\mathbf{p}%
_{j}-\mathbf{p}_{i}$. A maximum connectivity distance $d$ has been added to
the criterion in order to avoid unrealistic bonding.

\subsection{VA criterion}

By integrating the relative momenta weighted by the Maxwell distribution in
the region where the relative kinetic energy is lesser than minus the pair
potential, the momenta are eliminated and the VA bond conditional
probability density is obtained: \cite{Hill1,Coniglio2} 
\begin{equation}
P_{\text{VA}}(r_{i,j})=\left\{ 
\begin{array}{cc}
0 & v(r_{i,j})>0\text{ \ or \ }r_{i,j}>d \\ 
\gamma \lbrack 3/2,-v(r_{i,j})/k_{B}T]/\Gamma \lbrack 3/2] & v(r_{i,j})\leq 0%
\text{ \ and \ }r_{i,j}\leq d,%
\end{array}%
\right.  \label{4}
\end{equation}%
where $\Gamma \lbrack a]$ is the gamma function and $\gamma \lbrack a,x]$ is
the incomplete gamma function.

\subsection{Molecular dynamics calculations}

We consider a system of Lennard--Jones particles in a cubic box with
periodic boundary conditions in the $NVT$ ensemble and use a leap-frog
algorithm with velocity correction. \cite{Allen1} The time step is chosen as 
$\Delta t^{\ast }=\Delta t\sigma ^{-1}\sqrt{k_{B}T/(\varepsilon m)}=0.01$.
Quantities are averaged over $10^{3}$ configurations chosen every 100 $%
\Delta t$ after equilibration. A cut off distance equal to $2.5\sigma $ was
used in the pair potential.

In the VA case, for each pair of particles that satisfies $v(r_{i,j})\leq 0$
and $r_{i,j}\leq d$, we generate a random number $z$, between $0$ and $1$.
If $z<\gamma \lbrack 3/2,-v(r_{i,j})/k_{B}T]/\Gamma \lbrack 3/2]$ we
consider that the particles form a bonded-pair, otherwise we do not. Note
that this criterion can also be used in MC simulations because it does not
require information on the particle velocities. To identify the clusters
from the list of bonded-pairs we use the Stoddard's algorithm. \cite%
{Allen1,Stoddard1}

A system is said to be in a percolated state if a cluster that spans the
replicas is present 50 percent of the time. \cite{Seaton1} It is known that
this criterion yields results that are marginally affected by finite size
effects. \cite{Lee1} Then, a percolation transition curve, which separates
the percolated from the non-percolated states of the system, can be drawn
above the coexistence curve in the $T-\rho $ phase diagram.

In Fig. 1, the percolation loci for HE and VA connectivity criteria are
presented for $d=3\sigma $. These simulations were performed with $N=1372$
particles. The gas--liquid coexistence curve obtained by Panagiotopoulos 
\cite{Panagio1} and the MC liquid--solid coexistence curve of Hansen and
Verlet \cite{Hansen1} are also shown. The MD results for the HE criterion
are similar to those obtained by Campi \textit{et al.} \cite{Campi1} These
authors consider that the system is on the percolation line if the second
moment of the cluster size distribution that excludes the largest cluster $%
n^{\prime }(s)$ reaches its maximum. In Fig. 2 we show the dependence of the
calculated percolation density with system size and connectivity distance $d$%
. Extrapolation to infinite systems can be obtained by fitting a
straight-line to a plot of $\rho _{p}$ versus $L^{-1/\nu }$. \cite{Seaton1}
We have used the universal value of $\nu =0.88\pm 0.02$ reported by Gaunt
and Sykes \cite{Gaunt1} for three-dimensional systems. The error due to
finite size effects in the calculated value of $\rho _{p}$ for the $1372$%
-particle system is of $1.0\%$. This correction is smaller than the size of
our symbols in Fig. 1. Also from Fig. 2, we see that the effect of the
connectivity distance $d$ on $\rho _{p}$ is negligible for $d>2.5$.

As we can see from Fig. 1, the VA criterion is a very good approximation to
the full HE criterion as far as the percolation loci is concerned. In Ref.
I, we reported a VA percolation line that was located at much larger
densities, and concluded that the approximation was rather poor. The revised
results reported here show that this is not the case. The source of the
error in Ref. I comes from the way percolating clusters are detected
according to the Seaton--Glandt prescription. \cite{Seaton1} All clusters in
a given configuration are first identified by the Stoddard's algorithm, then
each cluster is analyzed separately to determine if its replicas are
connected with one another. Since the VA criterion implies the use of random
numbers to decide whether two particles are connected, the second step where
each separated cluster is analyzed for percolation needs to reuse the same
random numbers generated when it was first identified. This subtlety was
overseen in Ref. I, which led to the incorrect identification of actual
percolating clusters as disconnected replicas.

\section{Cluster pair correlation functions}

In the remainder of the paper we restrict our attention to the cluster pair
correlations for the HE and VA energetic criteria. We calculate them by
using the above mentioned integral equations and MD simulations. Thus, this
section will be devoted to pose the integral equations for the cluster
correlation functions $g_{\text{HE}}^{\dagger }(\mathbf{r}_{1},\mathbf{r}%
_{2})$ and $g_{\text{VA}}^{\dagger }(\mathbf{r}_{1},\mathbf{r}_{2})$ and
briefly discuss their solutions.

\subsection{VA criterion}

In order to study clustering in a system composed of $N$ classical particles
interacting via a pair potential $v(\mathbf{r}_{1},\mathbf{r}_{2})$, Hill
separated the Boltzmann factor $e(\mathbf{r}_{1},\mathbf{r}_{2})=\exp
[-\beta v(\mathbf{r}_{1},\mathbf{r}_{2})]$, into bonded $(\dagger )$ and
unbounded $(\ast )$ terms: \cite{Hill1} $e(\mathbf{r}_{1},\mathbf{r}%
_{2})=e^{\dagger }(\mathbf{r}_{1},\mathbf{r}_{2})+e^{\ast }(\mathbf{r}_{1},%
\mathbf{r}_{2})$. As usual $\beta =1/k_{B}T$, being $k_{B}$ the Boltzmann
constant. Since $e^{\dagger }(\mathbf{r}_{1},\mathbf{r}_{2})$ represents the
basic probability density of finding two particles bonded and at positions $%
\mathbf{r}_{1}$ and $\mathbf{r}_{2}$, this separation yields a diagrammatic
expansion for the partition function in terms of \textquotedblleft
chemical\textquotedblright\ clusters. We express Hill's separation as follows

\begin{equation}
e^{\dagger}(\mathbf{r}_{1},\mathbf{r}_{2})=P(\mathbf{r}_{1},\mathbf{r}_{2})%
\exp[-\beta v(\mathbf{r}_{1},\mathbf{r}_{2})]  \label{5}
\end{equation}

\begin{equation}
e^{\ast}(\mathbf{r}_{1},\mathbf{r}_{2})=[1-P(\mathbf{r}_{1},\mathbf{r}_{2})]%
\exp[-\beta v(\mathbf{r}_{1},\mathbf{r}_{2})]  \label{6}
\end{equation}
where $P(\mathbf{r}_{1},\mathbf{r}_{2})=P_{\text{VA}}(r_{1,2})$ is given by
Eq. (\ref{4}) in the case of the VA energetic criterion.

Fugacity and density expansions have been found, within Hill's formalism, by
Coniglio and co-workers \cite{Coniglio1} for the pair connectedness function 
$g^{\dagger }(\mathbf{r}_{1},\mathbf{r}_{2})\equiv g_{\text{VA}}^{\dagger }(%
\mathbf{r}_{1},\mathbf{r}_{2})$. As it was already mentioned, this function
is proportional to the joint probability density of finding two particles
belonging to the same cluster and at positions $\mathbf{r}_{1}$ and $\mathbf{%
r}_{2}$, respectively. Moreover, by collecting nodal and non-nodal diagrams
in these expansions an OZ-type relationship is obtained

\begin{equation}
g^{\dagger}(\mathbf{r}_{1},\mathbf{r}_{2})=c^{\dagger}(\mathbf{r}_{1},%
\mathbf{r}_{2})+\rho\int c^{\dagger}(\mathbf{r}_{1},\mathbf{r}%
_{3})g^{\dagger}(\mathbf{r}_{3},\mathbf{r}_{2})d\mathbf{r}_{3}.  \label{7}
\end{equation}

The function $c^{\dagger }(\mathbf{r}_{1},\mathbf{r}_{2})$ is the direct
pair connectedness function. By posing a closure relation, an integral
equation for $g^{\dagger }(\mathbf{r}_{1},\mathbf{r}_{2})$ is obtained.
Here, we use the more reliable closure available, \textit{i.e.} the PY-like
relation \cite{Coniglio1}

\begin{equation}
g^{\dagger }(r_{1,2})=[f^{\ast }(r_{1,2})+1][g^{\dagger
}(r_{1,2})-c^{\dagger }(r_{1,2})]+\exp [\beta
v(r_{1,2})]g(r_{1,2})f^{\dagger }(r_{1,2}).  \label{8}
\end{equation}%
In Eq. (\ref{8}), $f^{\ast }(r_{1,2})=e^{\ast }(r_{1,2})-1=\exp [-\beta
v(r_{1,2})][1-P_{\text{VA}}(r_{1,2})]-1$ is the unbound Mayer function and $%
g(r_{1,2})$ is the thermal pair distribution function (PDF).

In order to solve the integral equation given by Eqs. (\ref{7}) and (\ref{8}%
), for the Lennard--Jones potential we have implemented Labik's numerical
algorithm. \cite{Labik1}

\subsection{HE criterion}

\subsubsection{The integral equation}

We summarize here the basic theory that we have developed \cite{Pugnaloni4}
to describe the clustering and percolation for clusters whose bond
definition depends on the positions and momenta of the two particles under
consideration.

For a system of $N$ classical particles that interact through a pair
potential $v(\mathbf{r}_{i},\mathbf{r}_{j})$, we define a density
correlation function $\rho(\mathbf{r}_{1},\mathbf{r}_{2},\mathbf{p}_{1},%
\mathbf{p}_{2})$ that is $N(N-1)$ times the probability density of finding
two particles at the phase space configurations $(\mathbf{r}_{1}$, $\mathbf{p%
}_{1})$ and $(\mathbf{r}_{2}$, $\mathbf{p}_{2})$ respectively:

\begin{align}
\rho (\mathbf{r}_{1},\mathbf{r}_{2},\mathbf{p}_{1},\mathbf{p}_{2})& =\frac{%
N(N-1)}{h^{3N}N!Q(N,V,T)}  \notag \\
& \times \int \prod_{i=1}^{N}\exp [-\beta \frac{\mathbf{p}_{i}^{2}}{2m}%
]\prod_{i=1}^{N}\prod_{j>i}^{N}\exp [-\beta v(\mathbf{r}_{i},\mathbf{r}%
_{j})]d\mathbf{r}^{N-2}d\mathbf{p}^{N-2}.  \label{9}
\end{align}%
Here $h$ is Planck's constant and $Q(N,V,T)$ the canonical partition
function of the system. Then, in the same spirit of Hill and Coniglio 
\textit{et al.,} \cite{Hill1,Coniglio1} we separate $\exp [-\beta v(\mathbf{r%
}_{i},\mathbf{r}_{j})]$ into connecting and blocking parts,

\begin{equation}
\exp [-\beta v(\mathbf{r}_{i},\mathbf{r}_{j})]=f^{\dagger }(\mathbf{r}_{i},%
\mathbf{r}_{j},\mathbf{p}_{i},\mathbf{p}_{j})+f^{\ast }(\mathbf{r}_{i},%
\mathbf{r}_{j},\mathbf{p}_{i},\mathbf{p}_{j})+1.  \label{10}
\end{equation}%
Here $f^{\dagger }(\mathbf{r}_{i},\mathbf{r}_{j},\mathbf{p}_{i},\mathbf{p}%
_{j})$ represents the basic probability density that two particles in
configuration $(\mathbf{r}_{i},\mathbf{r}_{j},\mathbf{p}_{i},\mathbf{p}_{j})$
are bonded. We will sometimes use the shorthand notation $f^{\gamma }(%
\mathbf{r}_{i},\mathbf{r}_{j},\mathbf{p}_{i},\mathbf{p}_{j})\equiv
f_{i,j}^{\gamma }$, where $\gamma $ can be either $\dagger $ or $\ast $.
Substitution of Eq. (\ref{10}) in Eq. (\ref{9}) yields 
\begin{align}
\rho (\mathbf{r}_{1},\mathbf{r}_{2},\mathbf{p}_{1},\mathbf{p}_{2})& =\frac{%
N(N-1)}{h^{3N}N!Q(N,V,T)}\exp [-\beta v(\mathbf{r}_{1},\mathbf{r}_{2})] 
\notag \\
& \times \int \prod_{i=1}^{N}\exp [-\beta \frac{\mathbf{p}_{i}^{2}}{2m}]\sum
\{\prod f_{i,j}^{\dagger }f_{k,l}^{\ast }\}dr^{N-2}dp^{N-2},  \label{10b}
\end{align}%
where the sum is carried out over all possible arrangements of products of
functions $f_{i,j}^{\dagger }$ and $f_{k,l}^{\ast }$.

We note that the functions $f_{i,j}^{\dagger }$ and $f_{i,j}^{\ast }$ can
depend on the momenta as well as on the positions of the two particles, but
the sum of $f_{i,j}^{\dagger }$ and $f_{i,j}^{\ast }$ must be momentum
independent in order to conform to Eq. (\ref{10}). Except for this last
condition, the functions $f_{i,j}^{\dagger }$ and $f_{i,j}^{\ast }$ are
otherwise arbitrary for thermodynamic purposes. Of course, we choose them in
such a way that the desired definition of bonded particles for HE clusters
is achieved, \textit{i.e.},

\begin{equation}
f_{i,j}^{\dagger}=\exp[-\beta v(r_{i,j})]P(\mathbf{r}_{i,j},\mathbf{p}_{i,j})
\label{11}
\end{equation}

\begin{equation}
f_{i,j}^{\ast}=\exp[-\beta v(r_{i,j})][1-P(\mathbf{r}_{i,j},\mathbf{p}_{i,j})%
]-1  \label{12}
\end{equation}
where $P(\mathbf{r}_{i,j},\mathbf{p}_{i,j})=P_{\text{HE}}(\mathbf{r}_{i,j},%
\mathbf{p}_{i,j})$ is given in Eq. (\ref{3}).

Each term in the integrand of Eq. (\ref{10b}) can be represented as a
diagram consisting of two white $e_{1}$ and $e_{2}$ points, $N-2$ black $%
e_{i}$ points and some $f_{i,j}^{\dagger }$ and $f_{i,j}^{\ast }$
connections except between the white points. Here we take $e_{i}\equiv \exp
[-\beta \frac{\mathbf{p}_{i}^{2}}{2m}]$. White points are not integrated
over whereas black points are integrated over both their positions and
momenta. All the machinery normally used to handle standard diagrams in
classical liquid theory \cite{Hansen2} can now be extended to treat these
new type of diagrams. By following Coniglio's recipe to separate connecting
and blocking parts in the PDF, $g(\mathbf{r}_{1},\mathbf{r}_{2})=g^{\dagger
}(\mathbf{r}_{1},\mathbf{r}_{2},\mathbf{p}_{1},\mathbf{p}_{2})+g^{\ast }(%
\mathbf{r}_{1},\mathbf{r}_{2},\mathbf{p}_{1},\mathbf{p}_{2})$, we obtain an
OZ-like integral equation for $g^{\dagger }(\mathbf{r}_{1},\mathbf{r}_{2},%
\mathbf{p}_{1},\mathbf{p}_{2}),$%
\begin{align}
g^{\dagger }(\mathbf{r}_{1},\mathbf{r}_{2},\mathbf{p}_{1},\mathbf{p}_{2})&
=c^{\dagger }(\mathbf{r}_{1},\mathbf{r}_{2},\mathbf{p}_{1},\mathbf{p}_{2}) 
\notag \\
& +\int \rho (\mathbf{r}_{3},\mathbf{p}_{3})c^{\dagger }(\mathbf{r}_{1},%
\mathbf{r}_{3},\mathbf{p}_{1},\mathbf{p}_{3})g^{\dagger }(\mathbf{r}_{3},%
\mathbf{r}_{2},\mathbf{p}_{3},\mathbf{p}_{2})d\mathbf{r}_{3}d\mathbf{p}_{3}.
\label{13}
\end{align}%
Here $\rho (\mathbf{r}_{1},\mathbf{p}_{1})\rho (\mathbf{r}_{2},\mathbf{p}%
_{2})g^{\dagger }(\mathbf{r}_{1},\mathbf{r}_{2},\mathbf{p}_{1},\mathbf{p}%
_{2})$ is $N(N-1)$ times the joint probability density of finding two
particles at positions $\mathbf{r}_{1}$ and $\mathbf{r}_{2}$ with momenta $%
\mathbf{p}_{1}$ and $\mathbf{p}_{2}$, respectively, and belonging to the
same cluster, where the bonding criterion is given by Eqs. (\ref{11}), (\ref%
{12}) and (\ref{3}), while 
\begin{equation}
\rho (\mathbf{r}_{1},\mathbf{p}_{1})=\frac{1}{N-1}\int \rho (\mathbf{r}_{1},%
\mathbf{r}_{2},\mathbf{p}_{1},\mathbf{p}_{2})d\mathbf{r}_{2}d\mathbf{p}_{2}.
\label{14}
\end{equation}%
The function $c^{\dagger }(\mathbf{r}_{1},\mathbf{r}_{2},\mathbf{p}_{1},%
\mathbf{p}_{2})$ denotes the sum of all the non-nodal diagrams in the
diagrammatic expansion of $g^{\dagger }(\mathbf{r}_{1},\mathbf{r}_{2},%
\mathbf{p}_{1},\mathbf{p}_{2}).$ We recall here that a nodal diagram
contains at least one black point through which all paths between the two
white points pass. For a homogeneous system, we have

\begin{align}
g^{\dagger}(\mathbf{r}_{12},\mathbf{p}_{1},\mathbf{p}_{2}) & =c^{\dagger }(%
\mathbf{r}_{12},\mathbf{p}_{1},\mathbf{p}_{2})+\frac{\rho}{(2\pi
mk_{B}T)^{3/2}}  \notag \\
& \times\int\exp[-\beta\frac{p_{3}^{2}}{2m}]c^{\dagger}(\mathbf{r}_{13},%
\mathbf{p}_{1},\mathbf{p}_{3})g^{\dagger}(\mathbf{r}_{32},\mathbf{p}_{3},%
\mathbf{p}_{2})d\mathbf{r}_{3}d\mathbf{p}_{3}.  \label{15}
\end{align}

To obtain a closed integral equation with Eq. (\ref{13}) or Eq. (\ref{15}),
we need a closure relation between $g^{\dagger }(\mathbf{r}_{1},\mathbf{r}%
_{2},\mathbf{p}_{1},\mathbf{p}_{2})$ and $c^{\dagger }(\mathbf{r}_{1},%
\mathbf{r}_{2},\mathbf{p}_{1},\mathbf{p}_{2})$. Here we will use the PY
approximation $g(\mathbf{r}_{1},\mathbf{r}_{2})\exp [\beta v(\mathbf{r}_{1},%
\mathbf{r}_{2})]=1+N(\mathbf{r}_{1},\mathbf{r}_{2}),$ where the function $N(%
\mathbf{r}_{1},\mathbf{r}_{2})$ is the sum of the nodal diagrams in the
expansion of $g(\mathbf{r}_{1},\mathbf{r}_{2})$. Separation into connecting
and blocking parts, $g(\mathbf{r}_{1},\mathbf{r}_{2})=g^{\dagger }(\mathbf{r}%
_{1},\mathbf{r}_{2},\mathbf{p}_{1},\mathbf{p}_{2})+g^{\ast }(\mathbf{r}_{1},%
\mathbf{r}_{2},\mathbf{p}_{1},\mathbf{p}_{2})$ and $N(\mathbf{r}_{1},\mathbf{%
r}_{2})=N^{\dagger }(\mathbf{r}_{1},\mathbf{r}_{2},\mathbf{p}_{1},\mathbf{p}%
_{2})+N^{\ast }(\mathbf{r}_{1},\mathbf{r}_{2},\mathbf{p}_{1},\mathbf{p}_{2})$%
, yields

\begin{align}
g^{\dagger}(\mathbf{r}_{1},\mathbf{r}_{2},\mathbf{p}_{1},\mathbf{p}_{2}) &
=[f^{\ast}(\mathbf{r}_{1},\mathbf{r}_{2},\mathbf{p}_{1},\mathbf{p}%
_{2})+1][g^{\dagger}(\mathbf{r}_{1},\mathbf{r}_{2},\mathbf{p}_{1},\mathbf{p}%
_{2})-c^{\dagger}(\mathbf{r}_{1},\mathbf{r}_{2},\mathbf{p}_{1},\mathbf{p}%
_{2})]  \notag \\
& +\exp[\beta v(\mathbf{r}_{1},\mathbf{r}_{2})]g(\mathbf{r}_{1},\mathbf{r}%
_{2})f^{\dagger}(\mathbf{r}_{1},\mathbf{r}_{2},\mathbf{p}_{1},\mathbf{p}%
_{2}),  \label{16a}
\end{align}
or, for a homogeneous system,

\begin{align}
g^{\dagger}(\mathbf{r}_{12},\mathbf{p}_{1},\mathbf{p}_{2}) & =[f^{\ast }(%
\mathbf{r}_{12},\mathbf{p}_{1},\mathbf{p}_{2})+1][g^{\dagger}(\mathbf{r}%
_{12},\mathbf{p}_{1},\mathbf{p}_{2})-c^{\dagger}(\mathbf{r}_{12},\mathbf{p}%
_{1},\mathbf{p}_{2})]  \notag \\
& +\exp[\beta v(\mathbf{r}_{12}]g(\mathbf{r}_{12})f^{\dagger}(\mathbf{r}%
_{12},\mathbf{p}_{1},\mathbf{p}_{2}).  \label{16b}
\end{align}
Equation (\ref{13}) joined with Eq. (\ref{16a}) or Eq. (\ref{15}) joined
with Eq. (\ref{16b}) give a closed set of equations for $g^{\dagger}(\mathbf{%
r}_{1},\mathbf{r}_{2},\mathbf{p}_{1},\mathbf{p}_{2})$.

From the function $g_{\text{HE}}^{\dagger }(\mathbf{r}_{1},\mathbf{r}_{2},%
\mathbf{p}_{1},\mathbf{p}_{2})\equiv g^{\dagger }(\mathbf{r}_{1},\mathbf{r}%
_{2},\mathbf{p}_{1},\mathbf{p}_{2})$ we define the pair correlation function
for energetic clusters $g_{\text{HE}}^{\dagger }(\mathbf{r}_{1},\mathbf{r}%
_{2})$ according to Eq. (\ref{1}).

\subsubsection{Solution of the integral equation}

\subparagraph{Equivalence with an integral equation for polarizable fluids}

Our problem consists in solving Eq. (\ref{15}) for $g^{\dagger }(\mathbf{r}%
_{12},\mathbf{p}_{1},\mathbf{p}_{2})$ closed by the connectedness PY
relation (\ref{16b}) with $f^{\dagger }(\mathbf{r}_{i},\mathbf{r}_{j},%
\mathbf{p}_{i},\mathbf{p}_{j})$ and $f^{\ast }(\mathbf{r}_{i},\mathbf{r}_{j},%
\mathbf{p}_{i},\mathbf{p}_{j})$ given by Eqs. (\ref{11}) and (\ref{12}). In
the closure relation (\ref{16b}), $g(\mathbf{r}_{12})$ is the thermal PDF of
the system. In this work we take $g(\mathbf{r}_{12})$ from the solution of
the thermal OZ equation in the PY approximation. \cite{Hansen2}

An equation mathematically equivalent to Eq. (\ref{15}) has been previously
solved by Lado \cite{Lado1} in the study of nonpolar polarizable molecules.
Explicitly, the equation considered there, which is a generalized OZ
equation, relates the fluid total correlation function (TCF) $h(\mathbf{r}%
_{12},\mathbf{p}_{1},\mathbf{p}_{2})=g(\mathbf{r}_{12},\mathbf{p}_{1},%
\mathbf{p}_{2})-1$ (with $g(\mathbf{r}_{12},\mathbf{p}_{1},\mathbf{p}_{2})$
the PDF) and the direct correlation function (DCF) $c(\mathbf{r}_{12},%
\mathbf{p}_{1},\mathbf{p}_{2})$,

\begin{align}
h(\mathbf{r}_{12},\mathbf{p}_{1},\mathbf{p}_{2}) & =c(\mathbf{r}_{12},%
\mathbf{p}_{1},\mathbf{p}_{2})  \notag \\
& +\rho\int f\left( p_{3}\right) c(\mathbf{r}_{13},\mathbf{p}_{1},\mathbf{p}%
_{3})h(\mathbf{r}_{32},\mathbf{p}_{3},\mathbf{p}_{2})d\mathbf{r}_{3}d\mathbf{%
p}_{3},  \label{16c}
\end{align}
where $\mathbf{p}_{i}$ denotes the instantaneous dipolar moment induced on
molecule $i$ by the remaining molecules of the system. The function $f\left(
p\right) $ gives the instantaneous dipolar moment thermal distribution which
is assumed to have a Gaussian form

\begin{equation*}
f\left( p\right) =\frac{1}{\left( 2\pi\alpha/\beta\right) ^{3/2}}\exp\left( -%
\frac{\beta p^{2}}{2\alpha}\right) ,
\end{equation*}
where $\alpha$ is the effective polarizability of the molecules.

We observe that Eqs. (\ref{15}) and (\ref{16c}) are the same if we identify $%
h$ with $g^{\dagger }$, $c$ with $c^{\dagger }$, the induced dipolar moment $%
\mathbf{p}_{i}$ with the kinetic momentum $\mathbf{p}_{i}$ and the
polarizability $\alpha $ with the particle mass $m$. There are, however,
some differences between the connectivity problem and the
polarizable-molecule problem. The form of $f(p)$ does not need to be
Gaussian for polarizable molecules; moreover, $f(p)$ is coupled to the TCF.
Therefore, the value of the effective polarizability $\alpha $ depends on
the density and temperature of the system. In the connectivity problem,
however, the equivalent of $f(p)$, $\rho (\mathbf{r},\mathbf{p})/\rho $, is
intrinsically Gaussian and independent of the thermodynamic macrostate of
the system.

Another difference between the connectivity problem here and the problem
described by Lado is that our closure relation must be complemented with the
condition given by Eqs. (\ref{11}) and (\ref{12}). In addition, the closures
are different. Here we consider the connectedness version of PY whereas an 
\textit{almost} exact relation between DCF and TCF (van
Leeuwen--Groeneveld--De Boer \cite{vanLeeuwen1} exact relation with
approximate bridge function) is used by Lado. \cite{Lado1} Nevertheless,
these differences do not affect the general method of solution developed by
Lado and we can apply the same principle of expansions in orthogonal
functions.

Thus, following Lado, \cite{Lado1,Zarragoicoechea1} we start by reassigning
the unknown function to be the indirect correlation function

\begin{equation}
\gamma^{\dagger}(\mathbf{r}_{12},\mathbf{p}_{1},\mathbf{p}_{2})=g^{\dagger }(%
\mathbf{r}_{12},\mathbf{p}_{1},\mathbf{p}_{2})-c^{\dagger}(\mathbf{r}_{12},%
\mathbf{p}_{1},\mathbf{p}_{2}),  \label{17}
\end{equation}
rather than $g^{\dagger}(\mathbf{r}_{12},\mathbf{p}_{1},\mathbf{p}_{2})$,
and rewriting Eq. (\ref{15}) in Fourier representation,

\begin{align}
\tilde{\gamma}^{\dagger}(\mathbf{k},\mathbf{p}_{1},\mathbf{p}_{2}) & =\frac{%
\rho}{(2\pi mk_{B}T)^{3/2}}\dint d\mathbf{p}_{3}\exp[-\beta\frac{p_{3}^{2}}{%
2m}]  \notag \\
& \left[ \tilde{\gamma}^{\dagger}(\mathbf{k},\mathbf{p}_{1},\mathbf{p}_{3})+%
\tilde{c}^{\dagger}(\mathbf{k},\mathbf{p}_{1},\mathbf{p}_{3})\right] \tilde{c%
}^{\dagger}(\mathbf{k},\mathbf{p}_{3},\mathbf{p}_{2}).  \label{18}
\end{align}

The closure given by the PY relation [Eq. (\ref{16b})] together with the
conditions (\ref{11}), (\ref{12}) and (\ref{3}) yield

\begin{equation}
c^{\dagger}(\mathbf{r}_{12},\mathbf{p}_{1},\mathbf{p}_{2})=\left\{ 
\begin{array}{cc}
g(\mathbf{r}_{12})-\gamma^{\dagger}(\mathbf{r}_{12},\mathbf{p}_{1},\mathbf{p}%
_{2}) & \mathbf{p}_{1,2}^{2}/4m<-v(r_{12})\text{ \ and \ }r_{12}\leq d \\ 
\left( \exp[-\beta v(r_{12})]-1\right) \gamma^{\dagger}(\mathbf{r}_{12},%
\mathbf{p}_{1},\mathbf{p}_{2}) & \mathbf{p}_{i,j}^{2}/4m\geq -v(r_{12})\text{
\ or \ }r_{1,2}>d%
\end{array}
\right.  \label{19}
\end{equation}
The\ connectivity part of the PDF is then computed from $\gamma^{\dagger}$ as

\begin{equation}
g^{\dagger}(\mathbf{r}_{12},\mathbf{p}_{1},\mathbf{p}_{2})=\left\{ 
\begin{array}{cc}
g(\mathbf{r}_{12}) & \mathbf{p}_{1,2}^{2}/4m<-v(r_{12})\text{ \ and \ }%
r_{12}\leq d \\ 
\exp[-\beta v(r_{12})]\gamma^{\dagger}(\mathbf{r}_{12},\mathbf{p}_{1},%
\mathbf{p}_{2}) & \text{ }\mathbf{p}_{i,j}^{2}/4m\geq-v(r_{12})\text{ \ or \ 
}r_{1,2}>d%
\end{array}
\right.  \label{20}
\end{equation}
The Fourier transform in Eq. (\ref{18}) and its inverse are defined as

\begin{equation}
\tilde{f}\left( \mathbf{k}\right) =\dint d\mathbf{r}f\left( \mathbf{r}%
\right) e^{-i\mathbf{k.r}},  \label{21}
\end{equation}

\begin{equation}
f\left( \mathbf{r}\right) =\frac{1}{\left( 2\pi\right) ^{3}}\dint d\mathbf{k}%
\tilde{f}\left( \mathbf{k}\right) e^{i\mathbf{k.r}}.  \label{22}
\end{equation}

The standard method for solving Eqs. (\ref{18}) and (\ref{19}) is to
explicitly break out the angular dependence of all functions in the form of
expansions in spherical harmonics. \cite{Gray1}

\subparagraph{Expansion of the pair functions in orthogonal polynomials}

The essential point in the integral equation solution method \cite{Lado1} is
the expansion of all the pair functions, like $\gamma^{\dagger}(\mathbf{r}%
_{12},\mathbf{p}_{1},\mathbf{p}_{2})$, in terms of orthogonal polynomials.
First we expand

\begin{align}
\gamma^{\dagger}(\mathbf{r}_{12},\mathbf{p}_{1},\mathbf{p}_{2}) &
=\gamma^{\dagger}(r,p_{1},p_{2},\omega_{1},\omega_{2})  \notag \\
& =4\pi\tsum
\limits_{l_{1},l_{2},m}\gamma_{l_{1}l_{2}m}^{%
\dagger}(r,p_{1},p_{2})Y_{l_{1}m}\left( \omega _{1}\right) Y_{l_{2}\overline{%
m}}\left( \omega_{2}\right) ,  \label{23}
\end{align}
where $\omega_{1}$ and $\omega_{2}$ are the directions of the momenta $%
\mathbf{p}_{1}$ and $\mathbf{p}_{2}$, $\overline{m}=-m$, and $%
m=-l,-l+1,...,l $. In this and similar expressions, the vector $\mathbf{r}%
_{12}$ has been implicitly chosen as the $z$ direction in the specification
of the Euler angles $\omega=\left( \theta,\phi\right) $. The spherical
harmonics satisfy the orthogonality condition

\begin{equation}
\dint d\omega Y_{lm}\left( \omega\right)
Y_{l^{^{\prime}}m^{^{\prime}}}^{\ast }\left( \omega\right)
=\delta_{ll^{^{\prime}}}\delta_{mm^{^{\prime}}},  \label{24}
\end{equation}
so that the coefficients of the expansion (\ref{23}) are immediately
obtainable as 
\begin{equation}
\gamma_{l_{1}l_{2}m}^{\dagger}(r,p_{1},p_{2})=\frac{1}{4\pi}\dint
d\omega_{1}d\omega_{2}\gamma^{\dagger}(r,p_{1},p_{2},\omega_{1},\omega
_{2})Y_{l_{1}m}\left( \omega_{1}\right) Y_{l_{2}\overline{m}}^{\ast}\left(
\omega_{2}\right) .  \label{25}
\end{equation}

Similarly, we can break out the kinetic momentum dependence in the form of
expansions in polynomials of $p$,

\begin{equation}
\gamma_{l_{1}l_{2}m}^{\dagger}(r,p_{1},p_{2})=\tsum
\limits_{n_{1},n_{2}}\gamma_{l_{1}l_{2}m}^{\dagger\text{ }n_{1}n_{2}}\left(
r\right) Q_{n_{1}l_{1}}\left( p_{1}\right) Q_{n_{2}l_{2}}\left( p_{2}\right)
,  \label{26}
\end{equation}
which are constructed to be orthogonal with Gaussian weight function

\begin{equation}
f\left( p\right) =\frac{1}{(2\pi m/\beta)^{3/2}}\exp[-\beta p^{2}/2m],
\label{27}
\end{equation}
namely,

\begin{equation}
4\pi\dint \limits_{0}^{\infty}dpp^{2}f\left( p\right) Q_{nl}\left( p\right)
Q_{n^{^{\prime}}l}\left( p\right) =\delta_{nn^{^{\prime}}.}  \label{28}
\end{equation}
The coefficients of the expansion are then again obtainable by quadratures,

\begin{align}
\gamma_{l_{1}l_{2}m}^{\dagger\text{ }n_{1}n_{2}}\left( r\right) & =\dint
\limits_{0}^{\infty}dp_{1}dp_{2}\left[ 4\pi p_{1}^{2}f\left( p_{1}\right) %
\right] \left[ 4\pi p_{2}^{2}f\left( p_{2}\right) \right]  \notag \\
& \times\gamma_{l_{1}l_{2}m}^{\dagger}(r,p_{1},p_{2})Q_{n_{1}l_{1}}\left(
p_{1}\right) Q_{n_{2}l_{2}}\left( p_{2}\right) .  \label{29}
\end{align}
Given the Gaussian form of the weight function $f\left( p\right) $, the
associated polynomials are \cite{Morse1}

\begin{equation}
Q_{nl}\left( p\right) =\left[ \frac{\Gamma\left( \frac{1}{2}\left(
n-l\right) +1\right) \Gamma\left( \frac{3}{2}\right) }{\Gamma\left( \frac{1}{%
2}\left( n+l\right) +\frac{3}{2}\right) }\right] ^{1/2}\left( \frac{\beta
p^{2}}{2m}\right) ^{l/2}L_{\left( n-l\right) /2}^{l+1/2}\left( \frac{\beta
p^{2}}{2m}\right) ,  \label{30}
\end{equation}
where $L_{n}^{b}\left( t\right) $ are the associated Laguerre polynomials 
\cite{Abramowitz1} and $\Gamma\left( z\right) $ is the gamma function.

Accordingly, all the functions in $\mathbf{r}$-space are expanded in the form

\begin{equation}
\gamma^{\dagger}(\mathbf{r},\mathbf{p}_{1},\mathbf{p}_{2})=4\pi\dsum
\limits_{n_{1},\text{ }n_{2},\text{ }l_{1},\text{ }l_{2},\text{ }%
m}\gamma_{l_{1}l_{2}m}^{\dagger\text{ }n_{1}n_{2}}\left( r\right)
Q_{n_{1}l_{1}}\left( p_{1}\right) Q_{n_{2}l_{2}}\left( p_{2}\right)
Y_{l_{1}m}\left( \omega_{1}\right) Y_{l_{2}\overline{m}}\left( \omega
_{2}\right) ,  \label{31}
\end{equation}
where the $z$ axis is along $\mathbf{r}$ and the summation indices satisfy
the constraints

\begin{align}
n & =0,1,2,...,  \notag \\
l & =n,\text{ }n-2,\text{ }n-4,...,1\text{ or }0\text{,}  \label{32} \\
m & =0,\pm1,\pm2,...,\pm l.  \notag
\end{align}
The coefficients of Eq. (\ref{31}) can be obtained as

\begin{align}
\gamma_{l_{1}l_{2}m}^{\dagger\text{ }n_{1}n_{2}}\left( r\right) & =4\pi\dint
d\mathbf{p}_{1}d\mathbf{p}_{2}f\left( p_{1}\right) f\left( p_{2}\right)
\gamma^{\dagger}(\mathbf{r},\mathbf{p}_{1},\mathbf{p}_{2})  \notag \\
& \times Q_{n_{1}l_{1}}\left( p_{1}\right) Q_{n_{2}l_{2}}\left( p_{2}\right)
Y_{l_{1}m}^{\ast}\left( \omega_{1}\right) Y_{l_{2}\overline {m}%
}^{\ast}\left( \omega_{2}\right)  \label{33}
\end{align}
with $f\left( p\right) $ given by Eq. (\ref{27}). The complete
orthonormality condition is

\begin{equation}
4\pi\dint d\mathbf{p}f\left( p\right) Q_{nl_{{}}}\left( p\right)
Q_{n^{^{\prime}}l^{^{\prime}}}\left( p\right) Y_{lm}\left( \omega\right)
Y_{l^{^{\prime}}m^{^{\prime}}}^{\ast}\left( \omega\right)
=\delta_{nn^{^{\prime}}}\delta_{ll^{^{\prime}}}\delta_{mm^{^{\prime}}}.
\label{34}
\end{equation}

The functions in $\mathbf{k}$ can be expanded in a similar way. Setting the $%
z$ axis along $\mathbf{k}$, we write

\begin{equation}
\tilde{\gamma}^{\dagger }(\mathbf{k},\mathbf{p}_{1},\mathbf{p}_{2})=4\pi
\dsum\limits_{n_{1},\text{ }n_{2},\text{ }l_{1},\text{ }l_{2},\text{ }m}%
\tilde{\gamma}_{l_{1}l_{2}m}^{\dagger \text{ }n_{1}n_{2}}\left( k\right)
Q_{n_{1}l_{1}}\left( p_{1}\right) Q_{n_{2}l_{2}}\left( p_{2}\right)
Y_{l_{1}m}\left( \omega _{1}\right) Y_{l_{2}\overline{m}}\left( \omega
_{2}\right) .  \label{35}
\end{equation}%
However, the angles $\omega _{1}$ and $\omega _{2}$ are referred to
different axes in Eqs. (\ref{31}) and (\ref{35}), so that the coefficients
in these expansions are not themselves mutual Fourier transforms.

Introducing the expansion for $\tilde{\gamma}^{\dagger }(\mathbf{k},\mathbf{p%
}_{1},\mathbf{p}_{2})$ and the corresponding expansion for $\tilde{c}%
^{\dagger }(\mathbf{k},\mathbf{p}_{1},\mathbf{p}_{2})$, one finds that the
OZ-like equation in Fourier space [Eq. (\ref{18})] goes over into a set of
matrix equations for the respective coefficients,

\begin{equation}
\tilde{\gamma}_{l_{1}l_{2}m}^{\dagger\text{ }n_{1}n_{2}}\left( k\right)
=\left( -1\right) ^{m}\rho\tsum \limits_{n_{3},l_{3}}\left[ \tilde{\gamma}%
_{l_{1}l_{3}m}^{\dagger\text{ }n_{1}n_{3}}\left( k\right) +\tilde{c}%
_{l_{1}l_{3}m}^{\dagger\text{ }n_{1}n_{3}}\left( k\right) \right] \tilde{c}%
_{l_{3}l_{2}m}^{\dagger\text{ }n_{3}n_{2}}\left( k\right) .  \label{36}
\end{equation}

\subparagraph{Numerical procedure}

To obtain a numerical solution for the set of equations (\ref{15}) and (\ref%
{16b}) one needs the discrete versions of the expansion for $\gamma
^{\dagger }(\mathbf{r},\mathbf{p}_{1},\mathbf{p}_{2})$ [Eq. (\ref{31})] and
the quadratures for the coefficients $\gamma _{l_{1}l_{2}m}^{\dagger \text{ }%
n_{1}n_{2}}\left( r\right) $ [Eq. (\ref{33})]; these are

\begin{align}
\gamma^{\dagger}(r,i_{1},i_{2},k_{1},k_{2},j) & =4\pi\dsum \limits_{n_{1},%
\text{ }n_{2},\text{ }l_{1},\text{ }l_{2},\text{ }m}\gamma_{l_{1}l_{2}m}^{%
\dagger\text{ }n_{1}n_{2}}\left( r\right) Q_{n_{1}l_{1}}\left( i_{1}\right)
Q_{n_{2}l_{2}}\left( i_{2}\right)  \notag \\
& \times\mathcal{P}_{l_{1}m}\left( k_{1}\right) \mathcal{P}_{l_{2}\overline{m%
}}\left( k_{2}\right) \nu_{m}T_{m}\left( j\right)  \label{37}
\end{align}
and

\begin{align}
\gamma_{l_{1}l_{2}m}^{\dagger\text{ }n_{1}n_{2}}\left( r\right) & =\tsum
\limits_{i_{1},i_{2},k_{1},k_{2},j=1}^{N_{p}}w\left( i_{1}\right) w\left(
i_{2}\right) w\left( k_{1}\right) w\left( k_{2}\right) w\left( j\right)
\gamma^{\dagger}(r,i_{1},i_{2},k_{1},k_{2},j)  \notag \\
& \times Q_{n_{1}l_{1}}\left( i_{1}\right) Q_{n_{2}l_{2}}\left( i_{2}\right) 
\mathcal{P}_{l_{1}m}\left( k_{1}\right) \mathcal{P}_{l_{2}\overline{m}%
}\left( k_{2}\right) \left( -1\right) ^{m}T_{m}\left( j\right) .  \label{38}
\end{align}
In Eq. (\ref{37}), $\nu_{0}=1$ and $\nu_{m}=2$ for $m>0$. In Eq. (\ref{38}),
Gaussian quadratures are being used, with the argument $i$ standing for $%
t_{i}=\beta p_{i}^{2}/2m$, the $i$th root of $L_{N_{p}}^{1/2}\left( t\right) 
$, $k$ for $x_{k}=\cos\theta_{k}$, the $k$th root of $P_{N_{p}}\left(
x\right) $, and $j$ for $y_{j}=\cos\phi_{j}$, the $j$th root of $%
T_{N_{p}}\left( y\right) $, where $L_{N_{p}}^{1/2}\left( t\right) $, $%
P_{N_{p}}\left( x\right) ,$ and $T_{N_{p}}\left( y\right) $ are the
associated Laguerre, Legendre, and Chebyshev polynomials, respectively, all
of order $N_{p}$; here the associated Legendre functions $\mathcal{P}%
_{lm}(x) $ are normalized to 2. The $w$ are the corresponding Gaussian
weights,

\begin{equation}
w\left( i\right) =\left\{ t_{i}\left[ L_{N_{p}}^{1/2\prime}\left(
t_{i}\right) \right] ^{2}\right\} ^{-1},  \label{39}
\end{equation}

\begin{equation}
w\left( k\right) =\left\{ \left( 1-x_{k}^{2}\right) \left[
P_{N_{p}}^{\prime}\left( x_{k}\right) \right] ^{2}\right\} ^{-1},  \label{40}
\end{equation}

\begin{equation}
w\left( j\right) =N_{p}^{-1},  \label{41}
\end{equation}
where the prime denotes derivative.

The solution follows an iterative procedure. The preparatory stages of the
calculation consist of (i) computing the thermal PDF $g\left( r_{12}\right) $
for the Lennard--Jones fluid over a suitable mesh using the PY equation,
(ii) reducing the momentum space to the discrete set of points $\mathbf{p}%
_{i,k,j}\equiv \left( p_{i}\text{, }\theta _{k}\text{, }\phi _{j}\right) $ \
with $i,k,j=1,2,...,N_{p}$, and (iii) identifying the subset of states
---within all possible configurational states $(r_{12},\mathbf{p}_{1},%
\mathbf{p}_{2})$ of a pair of particles--- that correspond to a bonded pair.

We construct a logical array $\text{B}(r_{12},\mathbf{p}_{1;i,k,j},\mathbf{p}%
_{2;i,k,j})$ of dimension seven whose value is TRUE if the configurational
state of the pair of particles corresponds to a bonded state, \textit{i.e.,}
if $\ \mathbf{p}_{1,2}^{2}/4m<-v(r_{12})$ \ and $\ r_{12}\leq d$. If instead
this condition is not satisfied, then $\text{B}(r_{12},\mathbf{p}_{1;i,k,j},%
\mathbf{p}_{2;i,k,j})\text{ is FALSE.}$

The iterative solution of Eqs. (\ref{36}) and (\ref{19}) starts by guessing
the initial values of the coefficients $\gamma_{l_{1}l_{2}m}^{\dagger\text{ }%
n_{1}n_{2}}\left( r_{12}\right) $. Then, if $\text{B}(r_{12},\mathbf{p}%
_{1;i,k,j},\mathbf{p}_{2;i,k,j})\text{ is TRUE}$, following Eq. (\ref{20})
we take

\begin{equation}
g_{l_{1}l_{2}m}^{\dagger\text{ }n_{1}n_{2}}\left( r_{12}\right) =\QATOPD{\{}{%
.}{g\left( r_{12}\right) \text{ \ if \ }n_{1}=n_{2}=l_{1}=l_{2}=m=0}{0\text{
\ \ \ \ \ \ \ \ \ \ \ \ \ \ \ \ \ \ \ \ \ \ \ \ \ \ \ \ \ otherwise.\ \ \ \
\ \ \ \ \ \ \ \ \ \ }}  \label{42}
\end{equation}
If instead $\text{B}(r_{12},\mathbf{p}_{1;i,k,j},\mathbf{p}_{2;i,k,j})\text{
is FALSE}$ then, following Eq. (\ref{20}), we take

\begin{equation}
g_{l_{1}l_{2}m}^{\dagger \text{ }n_{1}n_{2}}\left( r_{12}\right) =\exp
[-\beta v(r_{12})]\gamma _{l_{1}l_{2}m}^{\dagger \text{ }n_{1}n_{2}}\left(
r_{12}\right) .  \label{43}
\end{equation}%
Knowing $g_{l_{1}l_{2}m}^{\dagger \text{ }n_{1}n_{2}}\left( r_{12}\right) $
and $\gamma _{l_{1}l_{2}m}^{\dagger \text{ }n_{1}n_{2}}\left( r_{12}\right) $
for all the mesh points and allowed indices, we can calculate [see Eqs. (\ref%
{17}) or (\ref{19})] 
\begin{equation}
c_{l_{1}l_{2}m}^{\dagger \text{ }n_{1}n_{2}}\left( r_{12}\right)
=g_{l_{1}l_{2}m}^{\dagger \text{ }n_{1}n_{2}}\left( r_{12}\right) -\gamma
_{l_{1}l_{2}m}^{\dagger \text{ }n_{1}n_{2}}\left( r_{12}\right) .  \label{44}
\end{equation}

We now need to transform the coefficients $c_{l_{1}l_{2}m}^{\dagger \text{ }%
n_{1}n_{2}}\left( r_{12}\right) $ in real space into coefficients $\tilde{c}%
_{l_{1}l_{2}m}^{\dagger \text{ }n_{1}n_{2}}\left( k\right) $ in Fourier
space. However, as we have mentioned, they are not themselves Fourier
transforms of each other. Thus, we have to assemble the complete function
first using the equation analogous to (\ref{37}) for $c^{\dagger
}(r,i_{1},i_{2},k_{1},k_{2},j)$ and then use a generalized fast-transform
algorithm \cite{Lado1} to calculate $\tilde{c}^{\dagger
}(k,i_{1},i_{2},k_{1},k_{2},j)$. Using the equation analogous to (\ref{38})
in $\mathbf{k}$-space we then have the coefficients $\tilde{c}%
_{l_{1}l_{2}m}^{\dagger \text{ }n_{1}n_{2}}\left( k\right) $ for the
complete set of indices and all the values of $k$ on an adequate mesh. The
coefficients $\tilde{\gamma}_{l_{1}l_{2}m}^{\dagger \text{ }%
n_{1}n_{2}}\left( k\right) $ are then easily calculated by using the OZ-like
equation in Fourier space [see Eq. (\ref{36})]. Again, we assemble the
complete function $\tilde{\gamma}^{\dagger }(k,i_{1},i_{2},k_{1},k_{2},j)$
[using the Fourier space version of Eq. (\ref{37})]. The inverse transform $%
\gamma ^{\dagger }(r_{12},i_{1},i_{2},k_{1},k_{2},j)$ is calculated with the
fast-transform algorithm and so new coefficients $\gamma
_{l_{1}l_{2}m}^{\dagger \text{ }n_{1}n_{2}}\left( r_{12}\right) $ [obtained
from Eq. (\ref{38})] are available to start again the iterative cycle. The
iterations end when convergence is reached, as measured by 
\begin{equation}
\left\vert \left[ \gamma _{l_{1}l_{2}m}^{\dagger \text{ }n_{1}n_{2}}\left(
r_{12}\right) \right] _{\left( s+1\right) \text{th iteration}}-\left[ \gamma
_{l_{1}l_{2}m}^{\dagger \text{ }n_{1}n_{2}}\left( r_{12}\right) \right] _{s%
\text{th iteration}}\right\vert <\epsilon  \label{45}
\end{equation}%
for the complete set of indices. The tolerance $\epsilon $ is set to 0.0001.

The pair correlation function for an energetic cluster [see Eq. (\ref{1})]
is finally given by

\begin{equation}
g_{\text{HE}}^{\dagger }(r_{12})=g_{000}^{\dagger \text{ }00}\left(
r_{12}\right) ,  \label{46}
\end{equation}%
where the orthonormality condition [see Eq. (\ref{34})] has been used.

\section{Results}

Firstly, as a complement to Fig. 1, we show in Fig. 3 the cluster pair
correlation functions $g_{\text{HE}}^{\dagger }(r_{12})$ and $g_{\text{VA}%
}^{\dagger }(r_{12})$ obtained from MD using $N=4000$ particles for ($\rho
^{\ast }$,$T^{\ast }$)\ = ($0.24$,$1.4$), ($\rho ^{\ast }$,$T^{\ast }$)\ = ($%
0.42$,$1.4$) and ($\rho ^{\ast }$,$T^{\ast }$)\ = ($0.429$,$1.4$) where $%
\rho ^{\ast }=\rho \sigma ^{3}$ and \ $T^{\ast }=k_{B}T/\varepsilon $. The
two last points correspond, respectively, to the VA and HE percolation loci
for $T^{\ast }=1.4$. The main peak in the cluster correlation functions is
higher for the VA criterion than for the HE criterion. This implies that
there is a larger tendency to consider as directly connected two neighbor
particles by this criterion. However, as it can be appreciated in Fig. 3, $%
g_{\text{VA}}^{\dagger }(r_{12})$ falls faster than $g_{\text{HE}}^{\dagger
}(r_{12})$ for larger $r$. This contrasting behavior can be understood by
analyzing the cluster size distribution function $n(s)$, which gives the
number of clusters in the system consisting of $s$ particles. Figure 4 shows 
$n(s)$ for the same three state points as Fig. 3. We can see here that the
VA criterion identifies a larger amount of clusters than the HE criterion up
to a certain size---which depends on the density. However the HE criterion
always identifies some clusters larger than the largest clusters identified
by the VA criterion. This means that the cluster correlation function is
more long ranged for the HE clusters.

Figs. 5 and 6 show the theoretical cluster correlation functions $g_{\text{HE%
}}^{\dagger }(r_{12})$ and $g_{\text{VA}}^{\dagger }(r_{12})$, calculated by
solving the corresponding integral equations [Eqs.(\ref{15}),(\ref{16b}) and
(\ref{7}),(\ref{8}), respectively] following the methods indicated in the
previous section, together with the corresponding MD simulation results. We
show the curves at temperatures $T^{\ast }=1.4$ and $3.0$ and densities $%
\rho ^{\ast }=0.24$ and $0.55$, respectively. This values are rather far
from the percolation loci since we have been unable so far to obtain
convergence of the numerical algorithms at higher densities when the HE
criterion is considered.

With generality, Figs. 5 and 6 say that the theoretical results follow quite
well the trends of the corresponding simulations for each criterion.

\section{Conclusions}

We have shown that, contrary to our previous results (Ref. I), the VA
energetic criterion is, in general, a good approximation to the full HE
criterion for estimating the percolation loci in a Lennard--Jones fluid.
However, the cluster correlation functions are somewhat different in the VA
case. We have obtained the cluster pair correlation functions for both
energetic criteria through the numerical integration of connectedness OZ
integral equations. In particular, we have used a generalization of the
integral equations that allows the implementation of the HE criterion. The
theoretical results agree rather well with the simulations.\bigskip

\begin{acknowledgments}
This work was supported by CONICET, CICPBA and UNLP (Argentina). We thank F.
Lado for providing the source code of his algorithm for the solution of the
nonpolar polarizable Lennard--Jones fluid.
\end{acknowledgments}

\bigskip\bigskip\

\newpage

\begin{center}
\textbf{Figure Captions}

\bigskip\ 
\end{center}

\textbf{Figure 1. }Coexistence and percolation curves for the Lennard--Jones
fluid. Open diamonds correspond to the gas--liquid coexistence.\cite%
{Panagio1} Open squares correspond to the fluid--solid coexistence.\cite%
{Hansen1} The percolation loci for the HE criterion (open circles) and for
the VA criterion (open triangles) are compared. Solid circles correspond to
the percolation loci for the HE criterion from Campi \textit{et al.}\cite%
{Campi1} Lines are only to guide the eye.

\textbf{Figure 2. }Percolation density for the HE criterion as a function of 
$L^{-1/\nu }$. $L$ is the simulation box length and $\nu =0.88$. \cite%
{Gaunt1} The largest system corresponds to $4000$ particles. The system size
used for the calculation of the percolation curves in Fig. 1 is indicated by
an arrow. The inset shows the percolation threshold as a function of $d$ for
the $1372$-particle system. The arrow shows the value of $d$ used in the
rest of the paper.

\textbf{Figure 3. }Connectedness correlation function for the HE criterion
(solid line) and for the VA criterion (dotted line) at $T^{\ast }=1.4$ for
various densities. The results for $\rho ^{\ast }=0.42$ and $0.429$
correspond to the percolation loci for the VA and the HE criterion,
respectively.

\textbf{Figure 4. }Cluster size distribution for the HE criterion (black
solid line) and for the VA criterion (red dotted line) at $T^{\ast }=1.4$
for the same densities as in Fig. 2.

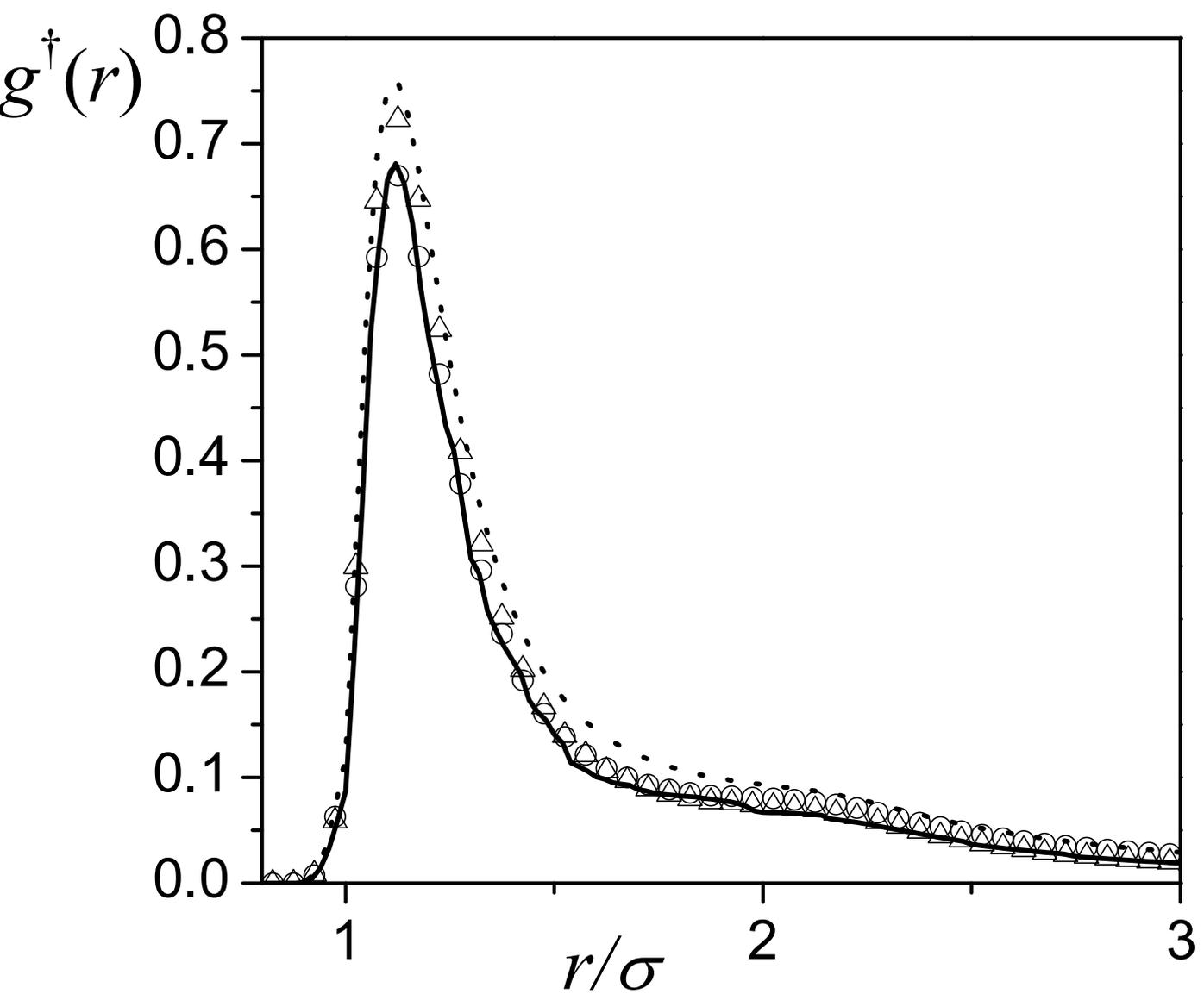
\textbf{Figure 5. }Connectedness correlation function for the HE criterion
(solid line and open circles) and for the VA criterion (dotted line and open
triangles) at $T^{\ast }=1.4$ and $\rho ^{\ast }=0.24$. Lines correspond to
the solution of the connectedness OZ equations closed with the connectedness
PY relation. Symbols correspond to MD.

\textbf{Figure 6. }Connectedness correlation function for the HE criterion
(solid line and open circles) and for the VA criterion (dotted line and open
triangles) at $T^{\ast }=3.0$ and $\rho ^{\ast }=0.55$.


\begin{thebibliography}{99}
\bibitem{comment1} For example, in the conductor--insulator transition of
water in oil microemulsions, two micelles are consider bonded if they can
share charge carriers. \cite{Chen1} However, in the identification of
bridges formed by grains within granular materials, two grains are bonded if
they stabilize each other. \cite{Pugnaloni1}
\end{thebibliography}
\end{document}